\documentclass[AMA,STIX1COL,doublespace]{WileyNJD-v2}

\usepackage{adjustbox}
\usepackage{threeparttable}
\usepackage{anyfontsize}
\usepackage{url}
\usepackage{bm}
\usepackage{amsmath}
\usepackage{mathtools}
\usepackage{xcolor}
\usepackage{graphicx}
\usepackage{rotating}
\usepackage{booktabs}

\articletype{Article Type}%

\received{26 April 2016}
\revised{6 June 2016}
\accepted{6 June 2016}

\begin{document}

%\title{Exposure measurement error correction in longitudinal studies with discrete outcomes\protect\thanks{This is an example for title footnote.}}
\title{Exposure measurement error correction in longitudinal studies with discrete outcomes}

\author[1]{Ce Yang}

\author[2]{Ning Zhang}

\author[1]{Jiaxuan Li}

\author[3]{Unnati V. Mehta}

\author[3,4,5]{Jaime E. Hart}

\author[6]{Donna Spiegelman}

\author[1,4,5,7]{Molin Wang*}

\authormark{Yang \textsc{et al}}

\address[1]{\orgdiv{Department of Epidemiology}, \orgname{Harvard T.H. Chan School of Public Health}, \orgaddress{\state{Boston, MA}, \country{USA}}}

\address[2]{\orgdiv{Department of Epidemiology}, \orgname{University of North Carolina}, \orgaddress{\state{Chapel Hill, NC}, \country{USA}}}

\address[3]{\orgdiv{Department of Environmental Health}, \orgname{Harvard T.H. Chan School of Public Health}, \orgaddress{\state{Boston, MA}, \country{USA}}}

\address[4]{\orgdiv{Channing Division of Network Medicine, Department of Medicine}, \orgname{Brigham and Women’s Hospital}, \orgaddress{\state{Boston, MA}, \country{USA}}}

\address[5]{\orgname{Harvard Medical School}, \orgaddress{\state{Boston, MA}, \country{USA}}}

\address[6]{\orgdiv{Department of Biostatistics}, \orgname{Yale School of Public Health}, \orgaddress{\state{New Haven, CT}, \country{USA}}}

\address[7]{\orgdiv{Department of Biostatistics}, \orgname{Harvard T.H. Chan School of Public Health}, \orgaddress{\state{Boston, MA}, \country{USA}}}

\corres{*Molin Wang, Department of Epidemiology, Harvard T.H. Chan School of Public Health, Boston, MA, USA. \\
\email{stmow@channing.harvard.edu}}

%\presentaddress{This is sample for present address text this is sample for present address text}

%%%%%%%%%%%%%%%%%%%%%%%%%%%%%%%%%%%%
%%%%%%%%%%%%%%%%%%%%%%%%%%%%%%%%%%%%
%%%%%%%%%%%%%%%%%%%%%%%%%%%%%%%%%%%%
\abstract[Summary]{Environmental epidemiologists are often interested in estimating the effect of time-varying functions of the exposure history on health outcomes.
However, the individual exposure measurements that constitute the history upon which an exposure history function is constructed are usually subject to measurement errors.
To obtain unbiased estimates of the effects of such mismeasured functions in longitudinal studies with discrete outcomes, a method applicable to the main study/validation study design is developed. 
Various estimation procedures are explored.
Simulation studies were conducted to assess its performance compared to standard analysis, and we found that the proposed method had good performance in terms of finite sample bias reduction and nominal coverage probability improvement. 
As an illustrative example, we applied the new method to a study of long-term exposure to PM$_{2.5}$, in relation to the occurrence of anxiety disorders in the Nurses’ Health Study II.
Failing to correct the error-prone exposure can lead to an underestimation of the chronic exposure effect of PM$_{2.5}$.}


%\jnlcitation{\cname{%
%\author{Williams K.},
%\author{B. Hoskins},
%\author{R. Lee},
%\author{G. Masato}, and
%\author{T. Woollings}} (\cyear{2016}),
%\ctitle{A regime analysis of Atlantic winter jet variability applied to evaluate HadGEM3-GC2}, \cjournal{Q.J.R. Meteorol. Soc.}, \cvol{2017;00:1--6}.}

\maketitle

%\footnotetext{\textbf{Abbreviations:} ANA, anti-nuclear antibodies; APC, antigen-presenting cells; IRF, interferon regulatory factor}

%%%%%%%%%%%%%%%%%%%%%%%%%%%%%%%
%%%%%%%%%%%%%%%%%%%%%%%%%%%%%%%
%%%%%%%%%%%%%%%%%%%%%%%%%%%%%%%
\section{Introduction}
\label{sec1}

\textcolor{black}{Modern epidemiologic studies often study longitudinal health endpoints from multiple follow-up questionnaires, for example, decline in cognitive function and change in body mass index. \citep{Weuve2012}}
While interest usually lies in the effect of a function of the exposure history, problems arise because the individual exposure measurements are typically error-prone and hence, will lead to biased estimates of the exposure effects on the health outcomes. \citep{Jacquez2011, Sheppard2012, Gerharz2013}

Anxiety disorder is a type of mental health condition that is characterized by disruptive fear, worry, and related behavioural disturbances. \citep{Kessler2009}
As the most common type of psychiatric disorder in the general population, anxiety disorders are associated with reduced productivity, increased medical care, absenteeism, and risk of suicide, leading to substantial monetary costs. \citep{Lepine2002,Greenberg1999}
Given the substantial personal and societal burden from anxiety and its persistent symptoms despite numerous available treatments \citep{Power2015, Lanouette2010}, it is essential to investigate population-level risk factors for anxiety disorders.
Here, we focus on studying air pollution, particularly chronic exposure to particulate matter less than 2.5 $\mu m$ in diameter (PM$_{2.5}$) in relation to the risk of developing anxiety disorders in the Nurses’ Health Study (NHS) II.
As an ongoing longitudinal cohort study of US female registered nurses, the NHS II has been collecting updated information biannually through mailed questionnaires that include questions on lifestyle and new diagnoses of health outcomes since 1989. 
About 116,429 participants aged 25 to 42 years old and living in 14 US states were enrolled in the study in 1989.
Although previous literature found a prospective association between PM$_{2.5}$ exposure and the development of major symptoms of anxiety \citep{Power2015, pun2017association}, the effect of chronic PM$_{2.5}$ exposure was not estimated accurately since the exposure data were mismeasured monthly exposure measurements constructed from a spatio-temporal model based on the nearest monitor to a person's residential addresses.
The correlation coefficient between such mismeasured exposure data and the gold standard, participants' personal exposure to PM$_{2.5}$ of ambient origin, was found to be 0.6 in Kioumourtzoglou et al. \cite{Kiou2014}
Therefore, exposure measurement error needs to be taken into account in order to produce a valid assessment of the effects of the history of the mis-measured exposures on the health outcomes of interest.

Although some statistical methods for longitudinal data analysis subject to covariate measurement error have been proposed \citep{Wang1998, Buonaccorsi2000, Shen2008, Liang2009, Zhang2018, Wong2018}, the existing methods are limited in their appropriateness for longitudinal environmental health studies.
First, as most methods apply only to a single exposure variable following the classical additive measurement error model for all participants at all time points, they are not useful for mismeasured exposure variables beyond this simple structure, nor to those that are functions of a mismeasured exposure history. \citep{Wang1998, Buonaccorsi2000, Shen2008, Liang2009}
Second, most existing methods are not designed for, and thus, not applicable to main study/validation study frameworks \citep{Zhang2018,Wong2018}, which are standard in environmental health research.
Most recently, Cai et al \citep{cai2023correcting} addressed these issues for longitudinal data analysis; however, this work only applies to continuous outcomes when the outcome model has a identity link function.
In this paper, we substantially generalize the work of Cai et al \citep{cai2023correcting} and develop a statistical method applicable to longitudinal health outcomes of arbitrary form, including the special case of discrete outcomes, that are linked to the linear predictor via a wide class of link functions.
\textcolor{black}{By leveraging validation studies, our method can validly estimate the true exposure using a general measurement error model based on the conditional expectation of the true given the error-prone measurements. 
This general formulation accommodates both strict Berkson error \citep{berkson1950there}, rare in practice and unsuitable for the data motivating our research, and strict classical errors. 
In the latter case, even if the data were generated under a classical error model, one can model and utilize the conditional distribution of the true given the error-prone measurements, although the classical error model is unsuitable for the data motivating our research and other similar applications.}
The resulting true exposure estimates facilitate the analysis of exposure history functions of interest in the main study framework.
Motivated by a study of PM$_{2.5}$ exposure in relation to anxiety disorders, our method will apply to any longitudinal study with repeated measures of outcomes and exposure variables that are functions of a mismeasured exposure history. 
\textcolor{black}{When there is only a single time point, longitudinal measurements of the exposure history will simplify to a single measurement, and our approach can still apply with a simplified estimation framework.}

The remainder of this paper is organized as follows.
We introduce our proposed method and outline the procedure for estimation and measurement error correction in Section \ref{sec2}.
In Section \ref{sec3}, we investigate the finite sample properties of our proposed estimator under main study/validation study design via simulation studies.
Particularly, we compare various estimation methods in the presence of an internal validation study.
In Section \ref{sec4}, we apply the method to analyze the motivating environmental health study which examines the effect of long-term exposure to PM$_{2.5}$ on the risk of developing anxiety disorders, as measured by the phobic anxiety subscale of the Crown-Crisp index. \citep{Crown1966} 
Section \ref{sec5} concludes with some final remarks.

%%%%%%%%%%%%%%%%%%%%%%%%%%%%%%%%
%%%%%%%%%%%%%%%%%%%%%%%%%%%%%%%%
%%%%%%%%%%%%%%%%%%%%%%%%%%%%%%%%
\section{Method for Measurement Error Correction}
\label{sec2}

%%%%%%%%%%%%%%%%%%%%%%%%%%%%%%
%%%%%%%%%%%%%%%%%%%%%%%%%%%%%%
\subsection{Notation}
\label{sec21}

We consider longitudinal main and validation studies of sizes $n_1$ and $n_2$, respectively.
Let $Y(t)$ denote the observed longitudinal outcome, $c(t)$, the true point exposure which is only available in the validation study, $C(t)$, the mismeasured point exposure which is available in both the main and validation studies, and $\textbf{W} (t)$, a vector of error-free covariates.
For individuals $i = 1, \ldots, n_1$ in the main study (MS), the observed data, $D_M$, include $m_i$ longitudinal measures 
$$D_M = \{Y_i (t_{i1}), \ldots, Y_i (t_{im_i}), C_i (t_{i1}), \ldots, C_i (t_{im_i}), \textbf{W}_i (t_{i1}), \ldots, \textbf{W}_i (t_{im_i}) \}$$
at time points $t_{ij}$, $j = 1, \ldots, m_i$.
When the true exposure is available in a representative subset of the study population, this subset forms an internal validation study (IVS).
Specifically, for individuals $i = n_1 + 1, \ldots, n_1 + n_2$ in the IVS, the data include
$$D_{IV} = \{Y_i (t_{i1}), \ldots, Y_i (t_{iv_i}), c_i (t_{i1}), \ldots, c_i (t_{iv_i}), C_i (t_{i1}), \ldots,
C_i (t_{iv_i}), \textbf{W}_i (t_{i1}), \ldots, \textbf{W}_i (t_{iv_i}) \}.$$
Here $v_i$ repeated measures are available for individual $i$.
On the other hand, an external validation study (EVS) typically recruits participants from a different group, so the data include
$$D_{EV} = \{c_i (t_{i1}), \ldots, c_i (t_{iv_i}), C_i (t_{i1}), \ldots, C_i (t_{iv_i}), \textbf{W}_i (t_{i1}), \ldots, \textbf{W}_i (t_{iv_i}) \}$$
for $i = n_1 + 1, \ldots, n_1 + n_2$.
We hereby note that transportability is assumed in such cases, i.e., the underlying relationship between the true and mismeasured exposure in the EVS is the same as that in the MS. \citep{rosner1990,Donna2000}
\textcolor{black}{While we allow individuals to have different measurement times in both main and validation studies, we also require that the true and mismeasured exposures are available at the same time points in the validation study.}

In epidemiologic studies, interest often lies in a function $h(\cdot)$ of the exposure history, for example, the moving average or the cumulative exposure \citep{Molin2016,Hart2015}.
Here, we let $s_i (t_{ij}) = \sum_{k=1}^{j-1} (t_{i(k+1)} - t_{ik}) c_i (t_{ik}) / (t_{ij} - t_{i1})$ denote the true cumulative average exposure of individual $i$ at time point $j$.
Similarly, let $S_i (t_{ij}) = \sum_{k=1}^{j-1} (t_{i(k+1)} - t_{ik}) C_i (t_{ik}) / (t_{ij} - t_{i1})$ denote the cumulative average exposure subject to measurement errors.
In what follows, we use tilde to indicate the history of a quantity up to a time point throughout the paper; for instance, $\tilde c_i (t_{ij}) = \{ c_i (t_{i1}), \ldots, c_i (t_{ij}) \}$ and $\tilde t_{ij} = \{ t_{i1}, \ldots, t_{ij} \}$ for individual $i$ at time point $j$.

%%%%%%%%%%%%%%%%%%%%%%%%%%%%%%
%%%%%%%%%%%%%%%%%%%%%%%%%%%%%%
\subsection{Conditional mean approximation under a regression model with a non-identity link function}
\label{sec22}

We assume that the regression model relates the true exposure and the covariates to the conditional mean of the outcomes through a link function $g(\cdot)$,
\begin{equation}
E[ Y_i(t_{ij}) | \tilde{c}_i(t_{ij}), \tilde{t}_{ij}, \tilde{\textbf{W}}_i(t_{ij}) ] = \mu_{ij} = g^{-1} \left( \textbf{X}_{ij}' \bm{\beta} \right),
\label{eq:ytrueexp}
\end{equation}
where $\textbf{X}_{ij} = \left( 1, h(\tilde{c}_i(t_{ij})), t_{ij}, h(\tilde{c}_i(t_{ij})) t_{ij}, \textbf{W}'_i (t_{ij}) \right)'$ for individual $i$ at time point $j$.
Throughout the paper, all vectors are column vectors.
The model is indexed by the parameter $\bm{\beta} = \left(\beta_0, \beta_1, \beta_2, \beta_3, \bm{\beta}_4' \right)'$, in which $\beta_3$, the parameter of interest, reflects the extent to which the trajectory of the outcome changes as the result of the exposure of interest over time.
By the law of total expectation,
\begin{align}
E[Y_i(t_{ij}) | \tilde{C}_i(t_{ij}), \tilde{t}_{ij}, \tilde{\textbf{W}}_i(t_{ij})] &= E_{c|C, \textbf{W}} \{ E[ Y_i(t_{ij}) | \tilde{c}_i(t_{ij}), \tilde{C}_i(t_{ij}), \tilde{t}_{ij}, \tilde{\textbf{W}}_i(t_{ij})] | \tilde{C}_i(t_{ij}), \tilde{t}_{ij}, \tilde{\textbf{W}}_i(t_{ij}) \} \nonumber \\
&= E_{c | C, \textbf{W}} \{ E[Y_i(t_{ij}) | \tilde{c}_i(t_{ij}), \tilde{t}_{ij}, \tilde{\textbf{W}}_i(t_{ij})] | \tilde{C}_i(t_{ij}), \tilde{t}_{ij}, \tilde{\textbf{W}}_i(t_{ij}) \} \nonumber \\
&= E_{c|C, \textbf{W}} \left[g^{-1} \left(\textbf{X}_{ij}' \bm{\beta} \right) \big | \tilde{C}_i(t_{ij}), \tilde{t}_{ij}, \tilde{\textbf{W}}_i (t_{ij}) \right], 
\label{S22doubleE}
\end{align}
where the second line follows from the surrogacy assumption. \citep{rosner1989}
In other words, conditional on the history of other covariates, $\tilde{\textbf{W}}_i$, the mismeasured exposure history $\tilde{C}_i$ does not contain additional information about $Y_i$ if the true exposure history $\tilde{c}_i$ is known.
Should we only observe the surrogates $\tilde{C}_i$, the conditional expectation in \eqref{S22doubleE} facilitates the analyses via the following generalized estimating equation (GEE) \citep{diggle2002},
\begin{align}
\bm{\psi_{\bm{\beta}}} &= \sum_{i=1}^N \bm{\omega}_i \left[ \textbf{Y}_i - E_{c|C, \textbf{W}} ( \bm{\mu}_i ) \right],
\label{S22GEE}
\end{align}
where $\bm{\omega}_i$ denotes an appropriate weight function of $C$ and $\textbf{W}$ but not $\textbf{Y}$, \textcolor{black}{and $\bm{\mu}_i = (\mu_{i1}, \ldots, \mu_{i m_i})'$ with $\mu_{ij}$ defined in \eqref{eq:ytrueexp}}. 
%$\textbf{Y}_i = \left(Y_i(t_{i1}), \ldots, Y_i(t_{im_i}) \right)'$, and $\bm{\mu}_i = (\mu_{i1}, \ldots, \mu_{i m_i})'$.
Then, given \eqref{S22doubleE}, $E[ \textbf{Y}_i - E_{c|C, \textbf{W}} ( \bm{\mu}_i )] = 0$ and \eqref{S22GEE} is unbiased.
Thus, we are left with finding a reasonably good approximation of the conditional expectation in \eqref{S22doubleE} that accommodates a general link function $g$.
This can be accomplished by performing a Taylor expansion of the function $g^{-1}(\cdot)$ around $ E \left[ \textbf{X}_{ij} \big | \tilde{C}_i(t_{ij}), \tilde{t}_{ij}, \tilde{\textbf{W}}_i(t_{ij}) \right]' \bm{\beta}$ following \eqref{S22doubleE} so that
{\normalsize
\begin{align}	
E[Y_i(t_{ij})|\tilde{C}_i(t_{ij}), \tilde{t}_{ij}, \tilde{\textbf{W}}_i(t_{ij})] &\approx E_{c|C, \textbf{W}} \left\{ g^{-1} \left( E \left[ \textbf{X}_{ij}' \bm{\beta} \big | \tilde{C}_i(t_{ij}), \tilde{t}_{ij}, \tilde{\textbf{W}}_i(t_{ij}) \right] \right) \right. \nonumber \\
&\left. \quad + \frac{d}{d (\textbf{X}_{ij}' \bm{\beta}) } g^{-1} \left( E \left[ \textbf{X}_{ij}' \bm{\beta} \big | \tilde{C}_i(t_{ij}), \tilde{t}_{ij}, \tilde{\textbf{W}}_i(t_{ij}) \right] \right) \left( \textbf{X}_{ij}' \bm{\beta} - E\left[\textbf{X}_{ij}' \bm{\beta} | \tilde{C}_i(t_{ij}), \tilde{t}_{ij}, \tilde{\textbf{W}}_i(t_{ij}) \right] \right) \right. \nonumber \\
&\left. \quad + \frac{1}{2!} \frac{d^2}{d (\textbf{X}_{ij}' \bm{\beta})^2} g^{-1} \left(E \left[ \textbf{X}_{ij}' \bm{\beta} \big | \tilde{C}_i(t_{ij}), \tilde{t}_{ij}, \tilde{\textbf{W}}_i(t_{ij}) \right] \right) \left(\textbf{X}_{ij}' \bm{\beta} - E\left[\textbf{X}_{ij}' \bm{\beta} | \tilde{C}_i(t_{ij}), \tilde{t}_{ij}, \tilde{\textbf{W}}_i(t_{ij}) \right] \right) \right. \nonumber \\ 
&\left. \quad \left(\textbf{X}_{ij}' \bm{\beta} - E \left[ \textbf{X}_{ij}' \bm{\beta} | \tilde{C}_i(t_{ij}), \tilde{t}_{ij}, \tilde{\textbf{W}}_i(t_{ij}) \right] \right)' \right\} \nonumber \\
&\approx g^{-1} \left( E \left[\textbf{X}_{ij} | \tilde{C}_i(t_{ij}), \tilde{t}_{ij}, \tilde{\textbf{W}}_i(t_{ij}) \right]' \bm{\beta} \right) + 0 \nonumber \\
&\quad + \frac{1}{2} \frac{d^2}{d (\textbf{X}_{ij}' \bm{\beta})^2} g^{-1} \left(E \left[ \textbf{X}_{ij} \big | \tilde{C}_i(t_{ij}), \tilde{t}_{ij}, \tilde{\textbf{W}}_i(t_{ij}) \right]' \bm{\beta} \right) \bm{\beta}' \text{Var} \left( \textbf{X}_{ij} | \tilde{C}_i(t_{ij}), \tilde{t}_{ij}, \tilde{\textbf{W}}_i(t_{ij}) \right) \bm{\beta}. \label{eq:ysurroexp}
\end{align}}
\textcolor{black}{The key to obtaining a valid numerical approximation is to assume that $\bm{\beta}$ is small (small effect) or $ \text{Var} (\textbf{X}_{ij} | \tilde{C}_i(t_{ij}), \tilde{t}_{ij}, \tilde{\textbf{W}}_i(t_{ij}))$ is small (small measurement error) so that higher order terms in the Taylor series expansion \eqref{eq:ysurroexp} can be ignored.
To empirically verify this assumption, one could estimate $\text{Var} (\textbf{X}_{ij} | \tilde{C}_i(t_{ij}), \tilde{t}_{ij}, \tilde{\textbf{W}}_i(t_{ij}))$, the residual variance in the measurement error model, in the validation study.
For cross-sectional study settings, the approximation has been found to work well when $\bm{\beta}' \text{Var} (\textbf{X}) \bm{\beta}$ is small based on simulation
studies of regression calibration for logistic regression. \citep{rosner1989,carroll1991semiparametric,neaton1992impact,kuha1994corrections}
Our simulation studies showed similar results in longitudinal settings, where the approximation worked well when $\bm{\beta}' \text{Var} (\textbf{X} | \tilde{C}, \tilde{t}, \tilde{W}) \bm{\beta}$ was less than 0.4.}
If the assumptions hold, then the following approximation
\begin{align}
E[Y_i(t_{ij}) | \tilde{C}_i(t_{ij}), \tilde{t}_{ij}, \tilde{\textbf{W}}_i(t_{ij})] &\approx g^{-1} \left( E \left[ \textbf{X}_{ij} | \tilde{C}_i(t_{ij}), \tilde{t}_{ij}, \tilde{\textbf{W}}_i(t_{ij}) \right]' \bm{\beta}  \right) \nonumber \\
&= g^{-1} \left( \beta_0 + E\left[h(\tilde{c}_i(t_{ij})) | \tilde{C}_i(t_{ij}), \tilde{t}_{ij}, \tilde{\textbf{W}}_i(t_{ij}) \right]' \beta_1 + \beta_2 t_{ij} \right. \nonumber \\ 
&\left. + E\left[ h(\tilde{c}_i(t_{ij})) | \tilde{C}_i(t_{ij}), \tilde{t}_{ij}, \tilde{\textbf{W}}_i(t_{ij}) \right]' \beta_3 t_{ij} + \tilde{\textbf{W}}_i' (t_{ij}) {\bm \beta}_4 \right)
\label{eq:ysurroapprox}
\end{align}
can replace the expectation $E ( \bm{\mu}_i )$ in GEE \eqref{S22GEE}.
Assume that the measurement error model is 
$$E \left[ c_i(t_{ik}) | C_i(t_{ik}), t_{ik}, \textbf{W}_i(t_{ik})\right] = \nu_{ik} = f\left(C_i(t_{ik}), t_{ik}, \textbf{W}_i(t_{ik}); \bm{\alpha} \right),$$
where $\bm{\alpha}$ is the unknown parameter, $i = n_1 + 1, \ldots, n_1 + n_2$, and $k = 1, \ldots, v_i$.
Suppose that the exposure history function of interest is the cumulative exposure average, i.e., $h(\tilde{c}_i(t_{ij})) = s_i (t_{ij})$, as in Weuve et al \cite{Weuve2012}.
The key component of the linear predictor in \eqref{eq:ysurroapprox} is then
\begin{align}
E \left[ s_{ij} | \tilde{C}_i(t_{ij}), \tilde{t}_{ij}, \tilde{\textbf{W}}_i(t_{ij}) \right] &= \frac{\sum_{k=1}^{j-1} (t_{i(k+1)}-t_{ik}) E\left[ c_i(t_{ik}) | \tilde{C}_i(t_{ij}), \tilde{t}_{ij}, \tilde{\textbf{W}}_i(t_{ij}) \right]}{t_{ij}-t_{i1}} \nonumber \\
&= \frac{\sum_{k=1}^{j-1} (t_{i(k+1)}-t_{ik}) E\left[ c_i(t_{ik}) | C_i(t_{ik}), t_{ik}, \textbf{W}_i(t_{ik}) \right]}{t_{ij}-t_{i1}} \nonumber \\
&= \frac{\sum_{k=1}^{j-1} (t_{i(k+1)}-t_{ik}) f \left(C_i(t_{ik}), t_{ik}, \textbf{W}_i(t_{ik}); \bm{\alpha} \right)}{t_{ij} - t_{i1}} \nonumber
\end{align}
under the localized error assumption \citep{zucker2008}, that conditional on $C_i(t_{ik})$, $t_{ik}$, and $\textbf{W}_i(t_{ik})$, variables at previous time points in the exposure history do not contain additional information about the conditional mean of $c_i(t_{ik})$.

%%%%%%%%%%%%%%%%%%%%%%%%%%%%%%
%%%%%%%%%%%%%%%%%%%%%%%%%%%%%%
\subsection{Estimation}
\label{sec23}

We now describe the procedure for estimating parameter $\bm{\theta} = (\bm{\alpha}', \bm{\beta}')'$ from estimating function $\bm{\psi} = (\bm{\psi_\alpha}', \bm{\psi_\beta}')'$, where $\bm{\psi_\alpha}$ and $\bm{\psi_\beta}$ refer to the estimating functions for the parameter, $\bm{\alpha}$, in the measurement error model, and parameters $\bm{\beta}$ and $\bm{\alpha}$, in the outcome model, respectively.
Solving the joint estimating equation $\bm{\psi} = \bm{0}$ consists of the following two steps.

First, we solve the GEE,
\begin{equation*}
\bm{\psi_{\bm{\alpha}}} = \sum_{i=n_1+1}^{n_1+n_2} \bm{\psi}_{i \bm{\alpha}} = \sum_{i=n_1+1}^{n_1+n_2} \left( \frac{\partial \bm{\nu}_i}{\partial \bm{\alpha}} \right)' \textbf{V}_{i}^{-1} \left( \textbf{c}_i - \bm{\nu}_i \right) = {\bm 0},
%\label{eq:eealphagee}
\end{equation*}
to obtain parameter estimate $\hat{\bm{\alpha}}$, which involves data from the validation study, where $\bm{\nu}_i = (\nu_{i1}, \ldots, \nu_{iv_i} )'$, $\textbf{V}_{i(v_i\times v_i)}$ is a working correlation matrix, and $\textbf{c}_i = (c_i(t_{i1}), \ldots, c_i(t_{iv_i}) )'$.
If we assume a linear measurement error model of the form
$$
c_i(t_{ik}) = \alpha_0 + \alpha_1 C_i(t_{ik}) + \alpha_2 t_{ik} + \alpha_3 C_i(t_{ik})t_{ik} + \textbf{W}_i' (t_{ik}) \bm{\alpha}_4 + \epsilon_i(t_{ik}),
%\label{eq:meeg}
$$
where $\epsilon_i$ denotes the residual term, the estimating equation simplifies to
\begin{equation}
\bm{\psi_{\bm{\alpha}}} = \sum_{i=n_1+1}^{n_1+n_2} \textbf{C}_i \textbf{V}_{i}^{-1} \left( \textbf{c}_i - \textbf{C}_i' \bm{\alpha} \right) = {\bm 0},
\label{eq:eealphageel}
\end{equation}
where $\textbf{C}_i = (\textbf{C}_{i1}, \ldots, \textbf{C}_{iv_i})$ with $\textbf{C}_{ik} = (1, C_i(t_{ik}), t_{ik}, C_i(t_{ik})t_{ik}, \textbf{W}'_i(t_{ik}) )'$ and $\bm{\alpha} = \left(\alpha_0, \alpha_1, \alpha_2, \alpha_3, \bm{\alpha}_4' \right)'$. 
\textcolor{black}{In particular, if there is only one measurement for each individual in the validation study, the estimation will simplify from a generalized estimating equation to an Ordinary Least Squares (OLS) regression.}
$\bm{\alpha}$ can be estimated using OLS regression,
\begin{equation}
\bm{\psi_{\bm{\alpha}}} = \sum_{i=n_1+1}^{n_1+n_2} \textbf{C}_{i1} \left( c_i - \textbf{C}_{i1}' \bm{\alpha} \right).
\label{eq:eealphaols}
\end{equation}

The second step estimates the true exposure from $\hat{\bm{\alpha}}$ as $\hat{c}_i(t_{ij}) = f \left( C_i(t_{ik}), t_{ik}, \textbf{W}_i(t_{ik}); \hat{\bm{\alpha}} \right)$
and computes the cumulative average exposure of the individuals as $\hat{s}_i(t_{i1}) = \hat{c}_i(t_{i1})$ and $\hat{s}_i(t_{ij}) = \sum_{k=1}^{j-1} \hat{c}_i(t_{ik}) (t_{i(k+1)}-t_{ik}) / (t_{ij}-t_{i1})$ for $j = 2, \ldots, m_i$.
Plugging $\hat{s}_i (t_{ij})$ into the mean model for $Y_i$, we solve the GEE 
\begin{equation}
\bm{\psi_{\bm{\beta}}} = \sum_{i=1}^{N} \bm{\psi}_{i \bm{\beta}} = \sum_{i=1}^{N} \left( \frac{\partial \hat{\bm{\mu}}_i}{\partial \bm{\beta}} \right)' \bm{\Sigma}_i^{-1} (\textbf{Y}_i - \hat{\bm{\mu}}_i) = \textbf{0}
\label{eq:eebetagee}
\end{equation}
to obtain the parameter estimate $\hat{\bm{\beta}}$,
where $N = n_1 + n_2$ for the MS/IVS design and $N = n_1$ for the MS/EVS design, $\hat{\bm{\mu}}_i = (\hat \mu_{i1}, \ldots, \hat \mu_{i m_i})'$, $\hat \mu_{ij} = g^{-1} (\hat{\textbf{X}}_{ij}' \bm{\beta}) = g^{-1} \left( (1, \hat{s}_i(t_{ij}), t_{ij}, \hat{s}_i(t_{ij}) t_{ij}, \textbf{W}_i' (t_{ij}) )' \bm{\beta} \right)$ for individual $i = 1, \ldots, N$ at time point $j = 1, \ldots, m_i$, and $\bm{\Sigma}_{i(m_i\times m_i)}$ is a working covariance matrix.

We note that there are alternatives for estimating $\bm{\beta}$ under the MS/IVS design. \citep{armstrong1985,Donna2005}
For example, one could consider using the true exposure  $c_i (t_{ij})$ for those in the IVS whenever they are available, i.e., replacing $\hat \mu_{ij}$ in \eqref{eq:eebetagee} as $\mu_{ij}$ for $i = n_1 + 1, \ldots, n_1 + n_2$ and $j = 1, \ldots, v_i$.
This estimator has a slightly different variance as shown in Appendix A of the Supplementary Materials.
Alternatively, one could use the IVS data alone to estimate $\bm{\beta}$ as $\hat{\bm{\beta}}_I$, followed by implementing the weighted average approach proposed by Spiegelman et al. \cite{Donna2001}
This leads to the inverse variance weighted average of the parameter estimates following from the MS and IVS analyses,
\begin{align}
    \hat{\bm{\beta}}_{IVW} = \frac{{\rm Var} (\hat{\bm{\beta}}_M)^{-1} \hat{\bm{\beta}}_M + {\rm Var} (\hat{\bm{\beta}}_I)^{-1} \hat{\bm{\beta}}_I }{ {\rm Var} (\hat{\bm{\beta}}_M)^{-1} + {\rm Var} (\hat{\bm{\beta}}_I)^{-1}},
    \label{S23betaivwa}
\end{align}
where $\hat{\bm{\beta}}_M$ is the solution to \eqref{eq:eebetagee} with $N = n_1$.
In this method, both the variance estimates of $\hat{\bm{\beta}}_M$ and $\hat{\bm{\beta}}_I$ are required;
such choice of weights gives the asymptotically most efficient combined estimate among all unbiased linear combinations of $\hat{\bm{\beta}}_M$ and $\hat{\bm{\beta}}_I$, since they are asymptotically uncorrelated. \citep{Donna2001}

\textcolor{black}{In summary, across the estimation procedures we discussed, we can obtain valid point and variance estimates under the following assumptions: (a) the surrogacy assumption \citep{rosner1989}, (b) the localized error assumption \citep{zucker2008}, (c) the small effect and/or small measurement error assumption for approximation, and (d) transportability for the MS/EVS design; that is, the measurement error model fitted in the validation study aligns with the main study data. Assumptions (b) and (c) are empirically verifiable; see Section \ref{sec42} for an example.}
See Section 2.4 and Appendix A of the Supplementary Materials for details.

%%%%%%%%%%%%%%%%%%%%%%%%%%%%%%
%%%%%%%%%%%%%%%%%%%%%%%%%%%%%%
\subsection{Variance estimation}
\label{sec24}

Following \eqref{eq:eealphageel}, $\hat{\bm{\alpha}}$ is fixed when substituted to $\bm{\psi_{\bm{\beta}}}$ and therefore, the solution to the joint estimating equation $\bm{\psi} = \bm{0}$ is equivalent to those resulting from the two steps.
The resulting estimator $\hat{\bm{\beta}}$ is consistent and asymptotically normal under both the MS/EVS and MS/IVS designs. \citep{Liang1986}
The asymptotic variance of $\hat{\bm{\beta}}$ is given by the sandwich formula, taking into account the variation induced by estimating $\hat{\bm{\alpha}}$ in the validation study.
Define square matrices $\textbf{B} \left( \bm{\theta} \right) = E \left( \frac{\partial \bm{\psi}}{\partial \bm{\theta}} \right)$ and $\textbf{A} \left( \bm{\theta}\right) = \text{Var} ( \bm{\psi} )$.
Then, the sandwich variance of $\hat{\bm{\theta}} = (\hat{\bm{\alpha}}', \hat{\bm{\beta}}')'$ is $\text{Var} (\hat{\bm{\theta}} ) = \textbf{B} \left( \bm{\theta} \right)^{-1} \textbf{A} \left( \bm{\theta} \right) \textbf{B} \left( \bm{\theta} \right)^{-1'}$.
The variance estimate of $\hat{\bm{\theta}}$ follows from evaluating estimates of $\textbf{A} \left( \bm{\theta}\right)$ and $\textbf{B} \left( \bm{\theta}\right)$ at $\hat{\bm{\theta}}$; $\text{Var} (\hat{\bm{\beta}} )$ is then estimated as the corresponding sub-matrix.
See Appendices A1 and A2 of the Supplementary Materials for details under both the MS/EVS and MS/IVS designs.
As for the inverse variance weighted estimator, we have $\widehat {\rm Var} (\hat{\bm{\beta}}_{IVW}) = (\widehat {\rm Var} (\hat{\bm{\beta}}_M)^{-1} + \widehat {\rm Var} (\hat{\bm{\beta}}_I)^{-1})^{-1}$.
This was shown in Spiegelman et al \cite{Donna2001} to improve efficiency over the original regression calibration estimator \cite{rosner1990,rosner1992} whenever the IVS is large enough to allow for a reasonable estimate of $\hat{\bm{\beta}}_I$.

%%%%%%%%%%%%%%%%%%%%%%%%%%%%%%%
%%%%%%%%%%%%%%%%%%%%%%%%%%%%%%%
%%%%%%%%%%%%%%%%%%%%%%%%%%%%%%%
\section{Simulation Studies}
\label{sec3}

In this section, we examine the finite sample behaviour of the estimators developed here, proposed under the MS/EVS and MS/IVS designs.
The performance of our new estimators was compared with that of the naive estimator which ignores measurement error.

%%%%%%%%%%%%%%%%%%%%%%%%%%%%%%
%%%%%%%%%%%%%%%%%%%%%%%%%%%%%%
\subsection{Simulation set-up}
\label{sec31}

\textcolor{black}{We considered simulation studies in which the main study sizes were $n_1 = 5000$ and 2000 and the validation study sizes were $n_2 = 500$ and 200.}
The number of simulation replicates was $500$.
For each individual in the validation study, we assumed a linear measurement error model 
\begin{equation}
c_i(t_{ij}) = \alpha_0 + \alpha_1 C_i(t_{ij}) + \alpha_2 t_{ij} + \alpha_3 C_i(t_{ij})t_{ij} + \alpha_4 W_i(t_{ij}) + \epsilon_i (t_{ij}),
\label{S31mem}
\end{equation}
at $v_i = 5$ time points.
Here, the first observed time point $t_{i1}$ followed a uniform distribution $U(0, 1)$.
The subsequent observations occurred at $t_{ij} = t_{i1} + j - 1$ for $j = 2, \ldots, 5$, so that the setting reflected a longitudinal study of a 5-year followup, with a variation in entry up to one year.
We considered both the surrogate exposure $C(t_{ij})$ and error-free covariates $W(t_{ij})$ to be scalars at each of the five time points.
They were generated from a multivariate normal distribution $\text{MVN}_{10 \times 10} (\textbf{0}, \bar{ \bf{V}})$ of the form
$\bar{ \bf{V}} = 
\begin{pmatrix} 
\bar{ \bf{V}}_C & \bar{ \bf{V}}_{C, W} \\
\bar{ \bf{V}}_{C, W} & \bar{ \bf{V}}_W
\end{pmatrix},$
\textcolor{black}{where we let $\bar{ \bf{V}}_C$ have a correlation structure of AR(1) with variance 1 and correlation coefficient 0.6, $\bar{ \bf{V}}_W$ a correlation structure of AR(1) with variance 1 and correlation coefficient 0.2, and $\bar{ \bf{V}}_{C, W}$ a diagonal correlation structure with diagonal elements 0.4 and off-diagonal elements 0.}
%where we let $\bar{ \bf{V}}_C$ follow an AR(1) structure with variance 1 and correlation coefficient 0.6, $\bar{ \bf{V}}_W$ an AR(1) structure with variance 1 and correlation coefficient 0.2, and $\bar{ \bf{V}}_{C, W}$ a correlation structure with diagonal elements 0.4 and off-diagonal elements 0.
The true exposure $c(t_{ij})$ was then generated from \eqref{S31mem} with $(\alpha_0, \alpha_1, \alpha_2, \alpha_3, \alpha_4) = (1.2, 0.6, 0.5, 0.4, 0.3)$.
The residual term $\epsilon_i(t_{ij})$ for $j = 1, \ldots, 5$ was assumed to be normal with zero mean and variance about 0.35 and 1.29 on average over time such that the correlation between $c(t_{ij})$ and $C(t_{ij})$ was controlled as 0.90 and 0.75, respectively.
We considered two cases, first, where only one single measurement and second, all five measurements of the true exposure were available in the validation study.
In the first case, the time point at which the true exposure measurement was available in the validation study was governed by a uniform random vector $U_i$ of length 5; specifically, the rank of the first element, following Liao et al. \cite{Liao2018}
The true and surrogate cumulative average exposure $s_i (t_{ij})$ and $S_i (t_{ij})$ then followed in the main and validation studies.

For each individual in the MS, we simulated Bernoulli responses $Y_i (t_{ij})$ at $m_i = 5$ time points, under the model
\begin{equation}
    \text{logit}[P \left( Y_i(t_{ij}) = 1 \right) | \tilde{c}_i (t_{ij}), \tilde{t}_{ij}, \tilde{W}_i(t_{ij})] = \beta_0 + \beta_1 s_i(t_{ij}) + \beta_2 t_{ij} + \beta_3 s_i(t_{ij}) t_{ij} + \beta_4 W_i(t_{ij}).
    \label{S31response}
\end{equation}
We generated the correlated binary responses, $Y_{i} (t_{ij})$, using an AR(1) correlation structure with a correlation coefficient of 0.1, where the marginal model was specified via \eqref{S31response} with $(\beta_0, \beta_1, \beta_2, \beta_3, \beta_4) = (-3, \log 1.2, 0.5, -\log 1.5, \log 1.1)$ or $(\beta_0, \beta_1, \beta_2, \beta_3, \beta_4) = (-3, \log 1.2, 0.5, -\log 1.1, \log 1.2)$ to reflect scenarios with disease prevalence about 5\% and 15\% on average over time, respectively.
Moreover, such specification of $\beta_3$ represented two cases with distinct strengths of the interaction between the cumulative average exposure and time.

%%%%%%%%%%%%%%%%%%%%%%%%%%%%%%%%%
%%%%%%%%%%%%%%%%%%%%%%%%%%%%%%%%%
\subsection{Simulation results}
\label{sec32}

Following the OLS regression in \eqref{eq:eealphaols} and the GEE in \eqref{eq:eealphageel}, $\hat s_i (t_{ij})$ was computed for $i = 1, \ldots, n_1$ and $j = 1, \ldots, m_i$ when one or multiple measurements of the true exposure were available in the validation study, respectively.
Such estimated $\hat s_i (t_{ij})$ facilitated the standard GEE analyses, 
% via the {\tt R} function {\tt geeglm};
where an AR(1) working correlation structure was specified.
% via the {\tt corstr} argument.
Under the MS/IVS design, $\hat s_i (t_{ij})$ was also computed for the validation subset $i = n_1 + 1, \ldots, n_1 + n_2$.
In the case when there was only a single measurement of the true exposure available in the IVS, $\hat{\bm{\beta}}_I$ in \eqref{S23betaivwa} was computed using a standard logistic regression where the single true exposure was treated as the cumulative average.
To assess the performance of our method, we computed the relative bias, $(\bar{\hat{\bm{\beta}}} - \bm{\beta}) \ /\bm{\beta}$, of our proposed estimators, where $\bar{\hat{\bm{\beta}}}$ refers to the averaged estimates over the simulation runs.
Recall that our interest lies in the estimation of $\beta_3$, the coefficient corresponding to the interaction term of the cumulative average exposure with time in the regression model \eqref{eq:ytrueexp}.
Furthermore, we computed the averaged standard errors (ASE) from the variance estimates developed in Section 2.4, together with the empirical standard errors (ESE).
To investigate the impact of exposure measurement error, the uncorrected estimates of $\bm{\beta}$ were computed for comparison.
Here, the measurement error was ignored in the GEE analyses, and $S_i (t_{ij})$ was used straightaway.
The ASEs of the uncorrected estimators also used the robust variance estimate formula.

Tables \ref{S32T1} and \ref{S32T2} display the results of estimators $\hat \beta_3$ under the MS/EVS and MS/IVS designs with various sample sizes when only one single measurement of the true exposure was available in the validation studies.
The ``Proposed" columns refer to the results of the estimators following from imputing the predicted exposure in both main and validation studies.
The ``Proposed - True" and ``Proposed - Inv" columns in Table \ref{S32T2} refer to those following from using the true exposure in the IVS whenever they were available and the inverse variance weighting method, respectively.
Our proposed methods following from \eqref{eq:eebetagee} showed consistently better results than the uncorrected analyses (the ``Uncorrected" columns) across all settings considered.
They took account of the measurement errors properly and led to estimators with modest relative biases and decent empirical coverage probabilities (CPs) in most cases.
Still, the performance of our proposed estimators tended to be better when the correlation between the true and surrogate exposures was higher and the interaction effect was small.
This is not surprising as the approximation in \eqref{eq:ysurroapprox} depends on the magnitude of both the measurement error and the value of parameter $\bm \beta$.
On the other hand, the inverse variance weighted estimator in \eqref{S23betaivwa} did not perform well when $\beta_3 = -\log 1.5$ as shown in the ``Proposed - Inv" column in Table \ref{S32T2}.
Here, the single true exposure was treated as the cumulative average exposure in the IVS, which was inaccurate and hence, led to poor performance especially when the IVS was large.

Additional results can be found in Tables 1 and 2 in Appendix B of the Supplementary Materials, including those from scenarios in which the true exposures were available at all five time points in the validation studies. 
The proposed methods following \eqref{eq:eebetagee} continued to perform better than the naive estimator.
Among the estimation methods under the MS/IVS design, estimators following \eqref{eq:eebetagee} had similar performance regardless of whether or not the true exposure was used in the IVS across the settings.
While the inverse variance weighted estimator in \eqref{S23betaivwa} did not do well when there was one single measurement of the true exposure in the IVS, this disappeared when the true exposure was available at all five time points in the IVS.
As shown in Table 2 in Appendix B of the Supplementary Materials, this estimator had the smallest relative biases in many of the cases.
Overall, we conclude that our proposed estimators had good finite sample performance and addressed the measurement error properly under both MS/EVS and MS/IVS designs.
When the true exposure was available at all time points in the IVS, the inverse variance weighting method performed well.
However, this method had limitations and was not suitable when the true exposure was missing at some assessments in the IVS.
In such cases, one should consider the estimators following \eqref{eq:eebetagee} instead.

%%%%%%%%%%%%%%%%%%%%%%%%%%%%%%%%
%%%%%%%%%%%%%%%%%%%%%%%%%%%%%%%%
\subsection{Robustness of the exposure measurement error correction procedure in longitudinal studies with discrete outcomes}
\label{sec33}

The results so far were obtained by adopting a measurement error model that correctly exploited the information of the surrogate exposure, time, and confounding covariates.
We next investigated the robustness of our proposed method to misspecification of the measurement error model.
Specifically, while the exposure data was generated via \eqref{S31mem}, we considered fitting a measurement error model in the validation study that failed to include the interaction between the surrogate exposure and time.
Such misspecification resulted in substantially biased estimators under both MS/EVS and MS/IVS designs.
See Table 3 in Appendix B of the Supplementary Materials for details.
Moreover, we considered employing various working correlation structures in the GEEs \eqref{eq:eealphageel} and \eqref{eq:eebetagee}.
In addition to the AR(1) structure, we adopted an independence and exchangeable working correlation structure for the longitudinal outcomes and obtained estimators with very similar relative biases and CPs.
See Table 4 in Appendix B of the Supplementary Materials for details.
Overall, our proposed method was robust to the misspecification of the outcome correlation structure in the main and validation study analyses.
However, the interaction term in \eqref{S31mem} was found to be crucial and empirically evident in our setting and it should be included in the measurement error model to ensure valid estimation and inference of the parameter of interest. 
In addition, when the main study model requires a time by exposure interaction term, the measurement error model also requires this.
See Appendix A3 of the Supplementary Materials for mathematical details.
\textcolor{black}{Lastly, we extended the simulation studies to settings where $(\beta_1, \beta_3) = (\log 1.2, -\log 2)$ and $\bm{\beta}' \text{Var} (\textbf{X} | \tilde{C}, \tilde{t}, \tilde{W}) \bm{\beta} = 0.4$, 0.6, 0.9. Although such a strong interaction effect may be rare in epidemiological applications, we use these settings to evaluate the finite sample  bias of our estimators when the small effect/small measurement error assumption does not hold. As expected, in our simulation studies, the relative biases of our $\hat \beta_3$-estimates were about $5\%$ to $6\%$ when $\bm{\beta}' \text{Var} (\textbf{X} | \tilde{C}, \tilde{t}, \tilde{W}) \bm{\beta} = 0.4$, and were $11\%$ to $14\%$ when $\bm{\beta}' \text{Var} (\textbf{X} | \tilde{C}, \tilde{t}, \tilde{W}) \bm{\beta}$ increased to 0.9.}
See Table 5 in Appendix B of the Supplementary Materials for details.

%%%%%%%%%%%%%%%%%%%%%%%%%%%%%%
%%%%%%%%%%%%%%%%%%%%%%%%%%%%%%
%%%%%%%%%%%%%%%%%%%%%%%%%%%%%%
\section{An Illustrative Example from the Nurses' Health Study II}
\label{sec4}

In this section, we apply our measurement error correction method to assess the impact of long-term exposure to PM$_{2.5}$ in relation to the development of anxiety disorders in the NHS II.
While the association between past PM$_{2.5}$ exposure and anxiety disorders was detected in literature using the surrogate exposure \citep{Power2015}, we aim to correct the bias in the estimation of the exposure history function effect induced by the measurement errors.

The rest of the section is structured as follows.
In Section \ref{sec41}, we provide with some details of the main and validation study data.
Method implementation is reported in Section \ref{sec42}.
The analyses and results are reported in Section \ref{sec43}.

%%%%%%%%%%%%%%%%%%%%%%%%%%%%%%
%%%%%%%%%%%%%%%%%%%%%%%%%%%%%%
\subsection{The study data}
\label{sec41}

%%%%%%%%%%%%%%%%%%%%%%%%%%%%%%
\subsubsection{Outcome assessment}

Anxiety symptoms in the NHS II were measured using the phobic anxiety subscale of the Crown-Crisp index (CCI). \citep{Crown1966}
The CCI is a standardized, self-rating questionnaire, and the phobic anxiety subscale contains eight questions about fearfulness and desire for avoidance of common situations or environments (having ``unreasonable fear of enclosed spaces", being ``scared of heights", disliking ``going out alone", feeling ``panic in crowds", feeling ``more relaxed indoors", feeling ``uneasy on buses or trains", tending to ``worry about getting incurable illness", worrying ``unduly when relatives are late coming home"), with two or three levels of possible responses each.
The total score ranges from 0 to 16, with higher scores indicating more anxiety. \citep{Crown1966}
Here, we used the subscale scores measured in 1993 and 2005.
For individuals with missing data on one or two questions out of eight, the total score was standardized by dividing by the fraction of the questions answered;
individuals who answered fewer than six questions were deleted. \citep{Brennan2009}
The CCI subscale scores were dichotomized at six, as prior research suggested that this cut-off represents a clinically relevant threshold such that those with a score of six points or higher have high symptoms of anxiety. \citep{McGrath2004,Okereke2012,Power2015}
These two repeated longitudinal binary outcomes were used in the analysis that follows.

%%%%%%%%%%%%%%%%%%%%%%%%%%%%%%%%%%%%%
\subsubsection{Exposure assessment}

The surrogate PM$_{2.5}$ exposure in the MS was obtained as follows.
A nationwide geographic information system (GIS)-based spatio-temporal model was used to estimate exposure to PM$_{2.5}$ for women residing in the United States.
See Yanosky et al \cite{yanosky2008spatio} for details of the exposure assessment models and their application in the evaluation of the impact of chronic PM or PM$_{2.5}$ exposures.
Briefly, data on PM$_{2.5}$ concentrations were obtained from the US Environmental Protection Agency’s (USEPA) Air Quality System (AQS) monitor nearest to each nurse's residence.
Such monitor data, together with additional GIS-based covariates, for instance, population density, distance to nearest roads, and weather variables \citep{Kiou2014,Hart2015}, were used to construct a generalized additive model to predict monthly PM$_{2.5}$ concentrations. 
Finally, the data were linked to each nurse by longitude and latitude of her last recorded address.

As for the true exposure, we used personal exposure to PM$_{2.5}$ of ambient origin available in an EVS including 426 observations of 274 individuals obtained from a number of panel studies performed in nine US cities between 1999 and 2002, the Multi-Ethnic Study of Atherosclerosis (MESA) study, and the Relationships of Indoor, Outdoor, and Personal Air (RIOPA) study. \citep{weisel2005relationships,turpin2007relationships,cohen2009approach,Kiou2014}
The true exposure was estimated by the sulfate ion ${\rm SO}_4^{2-}$, given that indoor sources were negligible and ${\rm SO}_4^{2-}$ had similar spatial homogeneity as PM$_{2.5}$. \citep{sarnat2002using,cai2023correcting}
Paired data on true and surrogate monthly exposure in the EVS allowed us to estimate the relationship between them through a measurement error model as in \eqref{S31mem}. 
About half of the EVS participants had true exposure measurements available at more than one time point.
By making the transportability assumption \citep{liao2011survival}, we used the measurement error model fitted in the EVS for measurement error correction in the following MS analysis that investigates the impact of 12-month moving average exposure to PM$_{2.5}$ prior to the anxiety assessments in 1993 and 2005 on the development of anxiety disorders.

%%%%%%%%%%%%%%%%%%%%%%%%%%%%%%%%%%%%%%%%
\subsubsection{Potential confounders}

We considered potential confounders that may affect the development of anxiety disorders following previous literature that studied the relation between exposure to fine particulate air pollution and prevalent anxiety. \citep{Power2015}
Potential individual-level confounders included age (continuous: years), marital status (binary: married vs not), husband's education level (binary: more than high school vs not), and calendar time of questionnaires return (categorical: six return periods).
Potential city-level confounders included percent of black race/ethnicity (continuous: zero to one) and median income (continuous: dollars per year).
We noted the inconsistency between the availability of the covariates in the MS and the EVS; 
this is typical under the MS/EVS design. \citep{Weuve2012}
We noted that risk factors for the outcome were not necessarily confounders and that in the context of air pollution exposure, which is determined weakly by individual characteristics if at all, confounding is typically minimal. \citep{weisskopf2017trade}
\textcolor{black}{Since interest lies in the association between anxiety disorders and chronic PM$_{2.5}$ exposure, we determined that a variable was not a confounder and could be removed from the model if the resulting parameter estimate, $\hat \beta_3$, following from the partial model did not change more than 10\% compared to the one following from the full model.}
\citep{maldonado1993simulation,cai2023correcting} 
Using this criterion, no variables other than age were included as confounders.

%%%%%%%%%%%%%%%%%%%%%%%%%%%%%%%%%%%%%%
%%%%%%%%%%%%%%%%%%%%%%%%%%%%%%%%%%%%%%
\subsection{Method implementation}
\label{sec42}

The sample sizes of the MS and EVS were 65158 and 274, respectively (Tables 1 and 2 in Appendix C of the Supplementary Materials).
First, we fitted a linear measurement error model of the form \eqref{S31mem} in the EVS, where the true monthly exposure was the dependent variable and the surrogate monthly exposure, age, and their interaction were the covariates.
Second, using the estimated regression coefficients from this measurement error model, we estimated the true exposures for each individual at each time point where the surrogate exposure was available in the MS. 
We then estimated the 12-month moving average for each individual at each time point using the estimated true exposures. 
Lastly, we fitted a logistic regression model with the dichotomized CCI subscale scores as the outcome and included the estimated 12-month moving average exposure to PM$_{2.5}$ prior to anxiety assessments, age, and their interaction as the covariates.
This logistic regression model can be written as
\begin{align*}
    \text{logit}[P \left( Y_i(t_{ij}) = 1 \right) | \tilde{c}_i (t_{ij}), \tilde{t}_{ij}] &= \beta_0 + \beta_1 \text{estimated 12-month moving average exposure}_i (t_{ij}) + \beta_2 \text{age}_{i} (t_{ij}) \\
    &+ \beta_3 \text{estimated 12-month moving average exposure}_i (t_{ij}) \text{age}_{i} (t_{ij}), 
\end{align*}
where $Y_i(t_{ij}) = 1$ if the CCI subscale score was at least six and $Y_i(t_{ij}) = 0$ if the CCI subscale score was less than six, for individual $i$ at time point $j$, as explained in Section \ref{sec41}.
The parameters in this model were estimated using GEE with an AR(1) working correlation structure.
The unstructured working covariance matrix was also explored, leading to nearly identical results.
The sandwich variance estimators in Section \ref{sec24} were then applied to obtain the variance and confidence interval estimates for the logistic regression parameters.  

\textcolor{black}{The surrogacy assumption was satisfied, as the exposure assessments occurred prior to disease onset.
The term $\bm{\beta}' \text{Var} (\textbf{X} | \tilde{C}, \tilde{t}, \tilde{W}) \bm{\beta}$ was calculated as 0.02 based on the estimates for $\bm{\beta}$ in Table \ref{S4T1}, which was sufficiently small for our proposed method to work.
We empirically verified the localized error assumption using the sequential ANOVA test \citep{cai2023correcting}, which indicated that, given the current surrogate exposure and age, adding the average of previous surrogate exposures to the model was not significantly associated with the current true exposure. 
Lastly, although not empirically verifiable in an MS/EVS design, it was reasonable to assume the transportability assumption, as the distributions of monthly spatial-temporal predicted PM$_{2.5}$ were similar between the MS and the EVS.}

%%%%%%%%%%%%%%%%%%%%%%%%%%%%%%%%%%%%%%
%%%%%%%%%%%%%%%%%%%%%%%%%%%%%%%%%%%%%%
\subsection{Results}
\label{sec43}

The estimated coefficient of the interaction term, $\hat{\beta}_3$, following our proposed method was 
$-5.81 \times 10^{-3}$ (95\% CI: [$-1.48 \times 10^{-2}$, $3.13 \times 10^{-3}$]) per $10 \mu g / m^3$ increase of prior 12-month moving average exposure to PM$_{2.5}$ per year, whereas using the surrogate exposure without measurement error correction led to 
$-5.51 \times 10^{-3}$ (95\% CI: [$-1.36 \times 10^{-2}$, $2.60 \times 10^{-3}$]);
see Table \ref{S4T1} for details.
%\textcolor{red}{Our result can be interpreted as, the odds ratio of developing anxiety disorders per 10$\mu g/m^3$ increment in the cumulative average exposure of PM$_{2.5}$ increases by 1\%  (95\% CI: [0\%, 2\%]) as the participant gets one year older.}
The difference between the estimates $\hat{\beta}_3$ indicated a stronger association between the prior 12-month moving average exposure to PM$_{2.5}$ and anxiety disorders upon measurement error correction.
In the uncorrected analysis, the odds ratio of developing anxiety disorders per $10 \mu g / m^3$ increment in prior 12-month moving average exposure to PM$_{2.5}$ was 1.12 (95\% CI: [1.04, 1.21]) at the age of 43, the median age of the MS participants.
Upon measurement error correction, however, the odds ratio became 1.14 (95\% CI: [1.05, 1.23]), per $10 \mu g / m^3$ increase in prior 12-month moving average exposure to PM$_{2.5}$.
While both odds ratios aligned with that in Power et al \cite{Power2015}, 1.15 (95\% CI: [1.06, 1.25]), failing to correct the error-prone exposure underestimated the effect estimates.
See Figure 1 in Appendix C of the Supplementary Materials for measurement error corrected and uncorrected odds ratios of developing anxiety disorders in the NHS II.
%See Figure 1 in Appendix C of the Supplementary Materials for the trend of the odds ratios of developing anxiety disorders for a given level of PM$_{2.5}$ exposure and baseline age using measurement error corrected and uncorrected estimates. The rate of increase in the odds ratios of developing anxiety disorders is higher when using the measurement error corrected estimates, compared to the estimated trajectory ignoring measurement error.
Finally, we comment that although we adjusted for age in both MS and EVS, the time scale used in the measurement error model does not have to be the same as that used in the outcome model for valid application of our proposed methods.
See Appendix A4 of the Supplementary Materials for mathematical details.

%%%%%%%%%%%%%%%%%%%%%%%%%%%%%%
%%%%%%%%%%%%%%%%%%%%%%%%%%%%%%
%%%%%%%%%%%%%%%%%%%%%%%%%%%%%%
\section{Discussion}
\label{sec5}

In this paper, we substantially generalized the work of Cai et al \citep{cai2023correcting} by developing a new method to correct the bias in the estimated effect of a exposure history function due to measurement error in longitudinal studies with outcomes of arbitrary form, including the important case of discrete outcomes.
Under the assumption of small effect size and/or small measurement error, our proposed approximation based on a Taylor series expansion provided a valid GEE analysis under both MS/EVS and MS/IVS designs.
Supported by the simulation studies, our proposed estimator performed well in finite sample bias reduction and improved the empirical coverage probabilities across the scenarios we studied.
In particular, choices of estimation methods under the MS/IVS design depended on the completeness of the true exposure measurements in the IVS data.
Investigating the effect of prior 12-month moving average exposure to PM$_{2.5}$ on the development of anxiety disorders in the NHS II, we found that the chronic exposure effect of PM$_{2.5}$ was underestimated when measurement error was ignored.

\textcolor{black}{Modeling the conditional expectation of the true exposure given the error-prone measurements is a Berkson-like approach, but it is not strictly Berkson, since the model $c = C + \epsilon$ rarely fits the data at hand, and certainly not in our study.
Our method is consistent with the regression calibration approach. \citep{rosner1989,carroll2006measurement} 
In our illustrative example, the measurement error process for air pollution exposure may be generated, in part, by a Berkson-like structure, where $c$ is influenced by $C$ and personal characteristics.
Similarly, in nutritional epidemiology, the underlying measurement error process may not follow the strict Berkson type. \citep{spiegelman1997regression, carroll2006measurement}
However, one can model the conditional expectation of $c$ given $C$ even if $C$ is produced from $c$ with some error in some way, including but not limited to $C = c + \epsilon$. 
In general, our method can be applied under the assumption that the model for the conditional expectation, $E(c|C, \textbf{W})$, is correctly specified, and that the regression coefficients of this model are consistent across the main and validation studies.
Note that if the measurement error model $E(c | C, \textbf{W}) = C$ holds at all the time points, the measurement errors will not introduce bias on the exposure-outcome associations. 
However, in practice, based on our experience, this is rarely, if ever, the case.}

\textcolor{black}{Our research was motivated by a previous study on air pollution exposure in relation to the development of anxiety disorders, where a strong association was reported. \citep{Power2015}
Other health outcomes, such as pulmonary symptoms, may also be closely related to air pollution exposure, and our method is applicable to any longitudinal study with repeated measures of outcomes and exposure variables that are functions of the exposure history, including those.}
In particular, the work is not limited to longitudinal discrete outcomes but can be applied whenever the outcome is related to the linear predictor via a non-identity link function in a regression model.
Future work would include investigating higher-order approximations to relax the small effect and measurement error assumptions we made.
One approach could be to include the higher order terms in the Taylor expansion \eqref{eq:ysurroexp}.
Measurement error correction in longitudinal studies following other analytic frameworks, such as maximum likelihood, is an important area of ongoing research as well.

%%%%%%%%%%%%%%%%%%%%%%%%%%%%%%%%%%%%%%%
%%%%%%%%%%%%%%%%%%%%%%%%%%%%%%%%%%%%%%%
%%%%%%%%%%%%%%%%%%%%%%%%%%%%%%%%%%%%%%%
%\backmatter

\section*{Acknowledgments}
This project was supported by the National Institute Health grants R01 DC017717, U01 CA176726 (NHS II) and the National Institute of Environmental Health Sciences grants 5R01ES026246, P30 ES000002.

%\subsection*{Author contributions}
%...

%\subsection*{Financial disclosure}
%None reported.

\subsection*{Conflict of interest}
The authors declare no potential conflict of interests.

%\section*{Supporting information}
%...

\subsection*{DATA AVAILABILITY STATEMENT}

The data that support the findings in this paper are not publicly available due to privacy or ethical restrictions.

%%%%%%%%%%%%%%%%%%%%%%%%%%%%%%%%%%%%%
%%%%%%%%%%%%%%%%%%%%%%%%%%%%%%%%%%%%%
%%%%%%%%%%%%%%%%%%%%%%%%%%%%%%%%%%%%%
%\nocite{*}% Show all bib entries - both cited and uncited; comment this line to view only cited bib entries;
\bibliography{wileyNJD-AMA}%

%%%%%%%%%%%%%%%%%%%%%%%%%%%%
%%%%%%%%%%%%%%%%%%%%%%%%%%%%
%%%%%%%%%%%%%%%%%%%%%%%%%%%
%\newpage

%%%%%%%%%%%%%%%%%%%%%%%%%%
\begin{table}[htbp]
\centering
\caption{Relative biases (RBias), average standard errors (ASE), empirical standard errors (ESE), and empirical coverage probabilities (CPs) of the 95\% confidence intervals of the estimator $\hat \beta_3$ following the proposed method and the uncorrected analyses under the MS/EVS design.
Only one measurement of the true exposure was available in the validation study.
The working correlation matrix was specified as AR(1) in the GEE analyses.
$n_1 = 5000/2000$, $n_2 = 500/200$, and $(\beta_1, \beta_3) = (\log 1.2, -\log 1.1) / (\log 1.2, -\log 1.5)$.
The correlation between the true and surrogate exposure was either 0.90 or 0.75.}
\label{S32T1}
\vspace{2mm}
\begin{tabular}{ccccccccccc}
\hline
\noalign{\medskip}
& & \multicolumn{4}{c}{Uncorrected} && \multicolumn{4}{c}{Proposed} \\
\cmidrule{3-6} \cmidrule{8-11} 
\noalign{\medskip}
$\beta_3$ & ${\rm Cor}(c, C)$ & RBias & ASE & ESE & CP && RBias & ASE & ESE & CP \\
\hline
	
\noalign{\medskip}
\multicolumn{11}{c}{$n_1 = 5000, n_2 = 500$} \\
\noalign{\medskip}
	
$-\log 1.1$ & 0.90 & 24.47\% & 0.015 & 0.015 & 0.66 && -0.05\% & 0.014 & 0.013 & 0.97 \\
& 0.75 & 24.28\% & 0.015 & 0.014 & 0.66 && -0.37\% & 0.014 & 0.013 & 0.97 \\
\noalign{\medskip}
$-\log 1.5$ & 0.90 & 26.11\% & 0.023 & 0.021 & 0.00 && -1.11\% & 0.030 & 0.021 & 0.99 \\
& 0.75 & 23.29\% & 0.023 & 0.021 & 0.01 && -4.95\% & 0.032 & 0.024 & 0.96 \\
	
\noalign{\medskip}
\multicolumn{11}{c}{$n_1 = 5000, n_2 = 200$} \\
\noalign{\medskip}
	
$-\log 1.1$ & 0.90 & 25.41\% & 0.015 & 0.015 & 0.66 && 0.28\% & 0.014 & 0.013 & 0.96 \\
& 0.75 & 25.37\% & 0.015 & 0.015 & 0.66 && -0.17\% & 0.014 & 0.014 & 0.97 \\
\noalign{\medskip}
$-\log 1.5$ & 0.90 & 26.23\% & 0.023 & 0.024 & 0.00 && -1.25\% & 0.031 & 0.026 & 0.98 \\
& 0.75 & 23.39\% & 0.023 & 0.023 & 0.01 && -5.57\% & 0.037 & 0.031 & 0.94 \\
	
\noalign{\medskip}
\multicolumn{11}{c}{$n_1 = 2000, n_2 = 500$} \\
\noalign{\medskip}
	
$-\log 1.1$ & 0.90 & 26.38\% & 0.024 & 0.025 & 0.79 && 0.88\% & 0.022 & 0.021 & 0.96 \\
& 0.75 & 26.24\% & 0.024 & 0.025 & 0.80 && 0.47\% & 0.022 & 0.021 & 0.95 \\
\noalign{\medskip}
$-\log 1.5$ & 0.90 & 26.93\% & 0.037 & 0.038 & 0.15 && -0.62\% & 0.046 & 0.037 & 0.99 \\
& 0.75 & 23.68\% & 0.036 & 0.038 & 0.22 && -4.98\% & 0.047 & 0.038 & 0.97 \\
	
\noalign{\medskip}
\multicolumn{11}{c}{$n_1 = 2000, n_2 = 200$} \\
\noalign{\medskip}
	
$-\log 1.1$ & 0.90 & 26.08\% & 0.024 & 0.024 & 0.80 && 1.09\% & 0.022 & 0.021 & 0.96 \\
& 0.75 & 26.46\% & 0.024 & 0.024 & 0.81 && 1.03\% & 0.023 & 0.021 & 0.96 \\
\noalign{\medskip}
$-\log 1.5$ & 0.90 & 26.33\% & 0.036 & 0.038 & 0.20 && -1.34\% & 0.047 & 0.038 & 0.98 \\
& 0.75 & 23.40\% & 0.036 & 0.037 & 0.30 && -5.69\% & 0.051 & 0.041 & 0.96 \\
\hline
\end{tabular}

\footnotesize{For a given MS sample size, relative biases improved as the validation study sample size increased.
In some cases, this was not observed for the estimate of the parameter of interest, $\beta_3$, but observed for the estimate of the parameter corresponding to the main effect of the cumulative average exposure, $\beta_1$.}
\end{table}
%%%%%%%%%%%%%%%%%%%%%%%%%%

%%%%%%%%%%%%%%%%%%%%%%%%%%
\begin{sidewaystable}[ht]
\centering
\caption{Relative biases (RBias), average standard errors (ASE), empirical standard errors (ESE), and empirical coverage probabilities (CPs) of the 95\% confidence intervals of the estimator $\hat \beta_3$ following the proposed method and the uncorrected analyses under the MS/IVS design.
Only one measurement of the true exposure was available in the validation study.
The working correlation matrix was specified as AR(1) in the GEE analyses.
$n_1 = 5000/2000$, $n_2 = 500/200$, and $(\beta_1, \beta_3) = (\log 1.2, -\log 1.1) / (\log 1.2, -\log 1.5)$.
The correlation between the true and surrogate exposure was either 0.90 or 0.75.}
\label{S32T2}
\vspace{2mm}
{\footnotesize
\begin{tabular}{ccccccccccccccccccccc}
\hline
\noalign{\medskip}
& & \multicolumn{4}{c}{Uncorrected} && \multicolumn{4}{c}{Proposed} && \multicolumn{4}{c}{Proposed - True} && \multicolumn{4}{c}{Proposed - Inv} \\
\cmidrule{3-6} \cmidrule{8-11} \cmidrule{13-16} \cmidrule{18-21}
\noalign{\medskip}
$\beta_3$ & ${\rm Cor}(c, C)$ & RBias & ASE & ESE & CP && RBias & ASE & ESE & CP && RBias & ASE & ESE & CP && RBias & ASE & ESE & CP \\
\hline
	
\noalign{\medskip}
\multicolumn{21}{c}{$n_1 = 5000, n_2 = 500$} \\
\noalign{\medskip}
	
$-\log 1.1$ & 0.90 & 24.01\% & 0.014 & 0.014 & 0.64 && -0.61\% & 0.013 & 0.012 & 0.96 && -0.60\% & 0.013 & 0.012 & 0.96 && -4.45\% & 0.014 & 0.013 & 0.95 \\
& 0.75 & 24.01\% & 0.014 & 0.014 & 0.64 && -0.81\% & 0.014 & 0.013 & 0.97 && -0.80\% & 0.014 & 0.013 & 0.96 && -5.97\% & 0.014 & 0.013 & 0.94 \\
\noalign{\medskip}
$-\log 1.5$ & 0.90 & 25.99\% & 0.022 & 0.022 & 0.00 && -1.20\% & 0.029 & 0.022 & 0.99 && -1.19\% & 0.029 & 0.022 & 0.98 && -10.37\% & 0.028 & 0.024 & 0.70 \\
& 0.75 & 23.09\% & 0.022 & 0.022 & 0.01 && -5.19\% & 0.031 & 0.025 & 0.94 && -5.11\% & 0.031 & 0.025 & 0.93 && -18.32\% & 0.030 & 0.026 & 0.28 \\
	
\noalign{\medskip}
\multicolumn{21}{c}{$n_1 = 5000, n_2 = 200$} \\
\noalign{\medskip}
	
$-\log 1.1$ & 0.90 & 25.50\% & 0.015 & 0.015 & 0.63 && 0.83\% & 0.014 & 0.013 & 0.95 && 0.84\% & 0.014 & 0.013 & 0.95 && -1.13\% & 0.014 & 0.013 & 0.95 \\
& 0.75 & 25.45\% & 0.015 & 0.015 & 0.61 && 0.34\% & 0.014 & 0.014 & 0.95 && 0.37\% & 0.014 & 0.014 & 0.95 && -2.22\% & 0.014 & 0.014 & 0.95 \\
\noalign{\medskip}
$-\log 1.5$ & 0.90 & 26.22\% & 0.023 & 0.025 & 0.00 && -1.59\% & 0.031 & 0.027 & 0.97 && -1.07\% & 0.031 & 0.027 & 0.97 && -5.12\% & 0.031 & 0.028 & 0.92 \\
& 0.75 & 23.29\% & 0.022 & 0.024 & 0.01 && -5.39\% & 0.036 & 0.032 & 0.93 && -5.33\% & 0.036 & 0.032 & 0.93 && -12.33\% & 0.035 & 0.034 & 0.70 \\
	
\noalign{\medskip}
\multicolumn{21}{c}{$n_1 = 2000, n_2 = 500$} \\
\noalign{\medskip}
	
$-\log 1.1$ & 0.90 & 25.70\% & 0.021 & 0.020 & 0.81 && 0.64\% & 0.020 & 0.018 & 0.97 && 0.66\% & 0.020 & 0.018 & 0.96 && -9.96\% & 0.021 & 0.019 & 0.94 \\
& 0.75 & 25.72\% & 0.021 & 0.020 & 0.81 && 0.27\% & 0.020 & 0.018 & 0.96 && 0.32\% & 0.020 & 0.018 & 0.96 && -12.95\% & 0.020 & 0.019 & 0.92 \\
\noalign{\medskip}
$-\log 1.5$ & 0.90 & 26.73\% & 0.033 & 0.031 & 0.08 && -0.67\% & 0.041 & 0.032 & 0.98 && -0.62\% & 0.041 & 0.032 & 0.99 && -20.38\% & 0.041 & 0.041 & 0.47 \\
& 0.75 & 23.38\% & 0.032 & 0.031 & 0.14 && -5.16\% & 0.043 & 0.034 & 0.96 && -4.95\% & 0.042 & 0.033 & 0.96 && -29.94\% & 0.040 & 0.042 & 0.15 \\
	
\noalign{\medskip}
\multicolumn{21}{c}{$n_1 = 2000, n_2 = 200$} \\
\noalign{\medskip}
	
$-\log 1.1$ & 0.90 & 26.11\% & 0.023 & 0.024 & 0.79 && 1.30\% & 0.021 & 0.021 & 0.95 && 1.30\% & 0.021 & 0.021 & 0.96 && -2.42\% & 0.022 & 0.021 & 0.95 \\
& 0.75 & 26.48\% & 0.023 & 0.024 & 0.79 && 1.11\% & 0.022 & 0.021 & 0.96 && 1.12\% & 0.022 & 0.021 & 0.96 && -3.82\% & 0.022 & 0.020 & 0.96 \\
\noalign{\medskip}
$-\log 1.5$ & 0.90 & 27.20\% & 0.035 & 0.035 & 0.12 && -0.43\% & 0.045 & 0.037 & 0.99 && -0.40\% & 0.045 & 0.037 & 0.99 && -9.19\% & 0.045 & 0.040 & 0.89 \\
& 0.75 & 24.04\% & 0.034 & 0.036 & 0.19 && -5.14\% & 0.049 & 0.041 & 0.96 && -5.02\% & 0.049 & 0.041 & 0.96 && -17.43\% & 0.047 & 0.045 & 0.67 \\
\hline
\end{tabular}
}

\footnotesize{``Proposed" refers to the method which uses measurement error corrected exposures for all participants;
``Proposed - True" refers to the method which uses true exposures whenever available in the IVS; 
``Proposed - Inv" refers to the method which leads to an inverse-variance weighted estimator.}
\end{sidewaystable}
%%%%%%%%%%%%%%%%%%%%%%%%%%

%%%%%%%%%%%%%%%%%%%%%%%%%%
\begin{sidewaystable}[ht]
\centering
\caption{Point estimate and 95\% confidence intervals (CIs) of the estimators following the proposed method and the uncorrected analyses studying the relationship between 12-month moving average exposure to PM$_{2.5}$ prior to anxiety assessments and anxiety disorders in the NHS II ($n_1 = 65158$ and $n_2 = 274$).
Personal PM$_{2.5}$ of ambient origin was treated as the true exposure and the estimated PM$_{2.5}$ from monitor data was treated as the surrogate exposure.
The estimates are in the scale of $10 \mu g / m^3$ of PM$_{2.5}$ per year.
The working correlation matrix was specified as AR(1) in the GEE analyses.}
\label{S4T1}
\vspace{2mm}

\begin{tabular}{ccccccc}
\hline
\noalign{\medskip}
&& \multicolumn{2}{c}{Uncorrected} && \multicolumn{2}{c}{Proposed} \\
\cmidrule{3-4} \cmidrule{6-7} 
\noalign{\medskip}
Parameter && $\hat \beta$ & $95\%$ CI && $\hat \beta$ & $95\%$ CI \\
\hline
	
\noalign{\medskip}

Prior 12-month moving average exposure && $3.55 \times 10^{-1}$ & [$-1.30 \times 10^{-3}$, $7.11 \times 10^{-1}$]  && $3.79 \times 10^{-1}$ & [$-1.22 \times 10^{-2}$, $7.70 \times 10^{-1}$]  \\
Age && $1.06 \times 10^{-1}$ & [$-2.61 \times 10^{-4}$, $2.13 \times 10^{-1}$]  && $8.64 \times 10^{-2}$ & [$5.44 \times 10^{-3}$, $1.67 \times 10^{-1}$]  \\
Age $\times$ exposure interaction && $-5.51 \times 10^{-3}$ & [$-1.36 \times 10^{-2}$, $2.60 \times 10^{-3}$]  && $-5.81 \times 10^{-3}$ & [$-1.48 \times 10^{-2}$, $3.13 \times 10^{-3}$]  \\

\hline
\end{tabular}
\end{sidewaystable}
%%%%%%%%%%%%%%%%%%%%%%%%%%

%%%%%%%%%%%%%%%%%%%%%%%%%%%%%%%%%%%%%%
%%%%%%%%%%%%%%%%%%%%%%%%%%%%%%%%%%%%%%
%%%%%%%%%%%%%%%%%%%%%%%%%%%%%%%%%%%%%%
%\section*{Author Biography}
%\begin{biography}{\includegraphics[width=66pt,height=86pt,draft]{empty}}{\textbf{Author Name.} This is sample author biography text.}
%\end{biography}

\end{document}